\documentclass[superscriptaddress,showpacs,preprintnumbers,amsmath,amssymb,twocolumn,prb]{revtex4-1}
\usepackage{graphicx}
\usepackage{verbatim} 
\usepackage[dvips]{color} 
\usepackage{epstopdf}

\usepackage{dcolumn}
\usepackage{bm}
\usepackage{upgreek}

\usepackage{color}

\begin{document}

\title{Single beam detection of optically driven spin dynamics\\ in CdTe/(Cd,Mg)Te quantum wells}


\author{F.~Saeed}
 	\affiliation{Experimentelle Physik 2, Technische Universit\"at Dortmund, 44221 Dortmund, Germany}
    \affiliation{Centre for Micro and Nano Devices, Department of Physics, COMSATS University Islamabad, Park Road, Islamabad 44000, Pakistan}
\author{M.~Kuhnert}
 	\affiliation{Experimentelle Physik 2, Technische Universit\"at Dortmund, 44221 Dortmund, Germany}
\author{I.~A.~Akimov}
 	\affiliation{Experimentelle Physik 2, Technische Universit\"at Dortmund, 44221 Dortmund, Germany}
 	\affiliation{Ioffe Institute, Russian Academy of Sciences, 194021 St. Petersburg, Russia}
\author{V.~L.~Korenev}
 	\affiliation{Ioffe Institute, Russian Academy of Sciences, 194021 St. Petersburg, Russia}
\author{G.~Karczewski}
\author{M.~Wiater}
\affiliation{Institute of Physics, Polish Academy of Sciences, PL-02668 Warsaw, Poland}
\author{T.~Wojtowicz}
\affiliation{Institute of Physics, Polish Academy of Sciences, PL-02668 Warsaw, Poland}
\affiliation{International Research Centre MagTop, PL-02668 Warsaw, Poland}
\author{A.~Ali}
\author{A.~S.~Bhatti}
\affiliation{Centre for Micro and Nano Devices, Department of Physics, COMSATS University Islamabad, Park Road, Islamabad 44000, Pakistan}
\author{D.~R.~Yakovlev}
\author{M.~Bayer}
 	\affiliation{Experimentelle Physik 2, Technische Universit\"at Dortmund, 44221 Dortmund, Germany}
 	\affiliation{Ioffe Institute, Russian Academy of Sciences, 194021 St. Petersburg, Russia}

\date{\today}

\begin{abstract}

We study optical pumping of resident electron spins under resonant excitation of trions in n-type CdTe/(Cd,Mg)Te quantum wells subject to a transverse magnetic field. In contrast to the comprehensively used time-resolved pump-probe techniques with polarimetric detection, we exploit here a single beam configuration in which the time-integrated intensity of the excitation laser light transmitted through the quantum wells is detected. The transmitted intensity reflects the bleaching of light absorption due to optical pumping of the resident electron spins and can be used to evaluate the Larmor precession frequency of the optically oriented carriers and their spin relaxation time. Application of the magnetic field leads to depolarization of the electron spin ensemble so that the Hanle effect is observed. Excitation with a periodic sequence of laser pulses leads to optical pumping in the rotating frame if the Larmor precession frequency is synchronized with the pulse repetition rate. This is manifested by the appearance of Hanle curves every 3.36 or 44.2~mT for pulse repetition rates of 75.8 or 999~MHz, respectively. From the experimental data we evaluate the $g$ factor of $|g|=1.61$ and the spin relaxation time of 14~ns for the optically pumped resident electrons, in agreement with previous time-resolved pump-probe studies.

\end{abstract}


\maketitle

\section{Introduction}

The non-equilibrium spin dynamics in semiconductor nanostructures attract particular interest due to possible applications in spintronic devices~\cite{Zutic-04}. In this regard, optical methods are considered to be promising for control and detection of electron spins~\cite{Awschalom-book}. One of the most common techniques for the evaluation of spin relaxation times of photoexcited as well as resident carriers in semiconductors is based on the Hanle effect, which is manifested by the depolarization of the photoluminescence under application of a transverse magnetic field~\cite{Dyakonov-book}. In the simplest case, the depolarization curve is described by a Lorentz profile with a half-width at half maximum $B_{1/2}=\hbar/g \mu_B T_s$, where $\hbar$ is the reduced Plank constant, $g$ is the Land\'{e} factor of the spin polarized carriers (e.g. conduction band electrons or valence band holes), $\mu_B$ is the Bohr magneton, and $T_s$ is the spin lifetime which is determined by the spin relaxation time $\tau_s$ and the lifetime $\tau$ of the optically oriented carriers as $T_s^{-1}=\tau^{-1}+\tau_s^{-1}$. However, the evaluation of the spin lifetime from Hanle curve requires precise knowledge of the $g$ factor.

Significant advancements in optical control of spin states and enhanced insight into the underlying dynamics have been achieved by means of time-resolved techniques such as pump-probe Faraday/Kerr rotation~\cite{Crooker-97,Harley-94}. Here, the use of pulsed laser sources arranged in a pump-probe setting allows one to resolve the pump-induced spin dynamics in time and to evaluate all required parameters such as the $g$ factor and the spin relaxation time~\cite{Awschalom-book,Dyakonov-book}. Especially interesting is the regime, in which the spin lifetime exceeds the pulse separation of the laser source. In this case, resonant spin amplification occurs if the Larmor precession frequency $\Omega_L=g\mu_BB/\hbar$ is a multiple of $2\pi f$, where $f$ is the laser pulse repetition rate and $B$ is the external magnetic field strength~\cite{Kikawa-98, Zhukov-07, Yugova-09, Korn-10, spin-inetria-15, steps-18}. However, time-resolved pump-probe Faraday or Kerr rotation measurements require a rather complex optical adjustment of the two beams, the mechanical delay line and the polarization sensitive detection.

\begin{figure*}[htpb]
\centering
\includegraphics[width=0.7\linewidth]{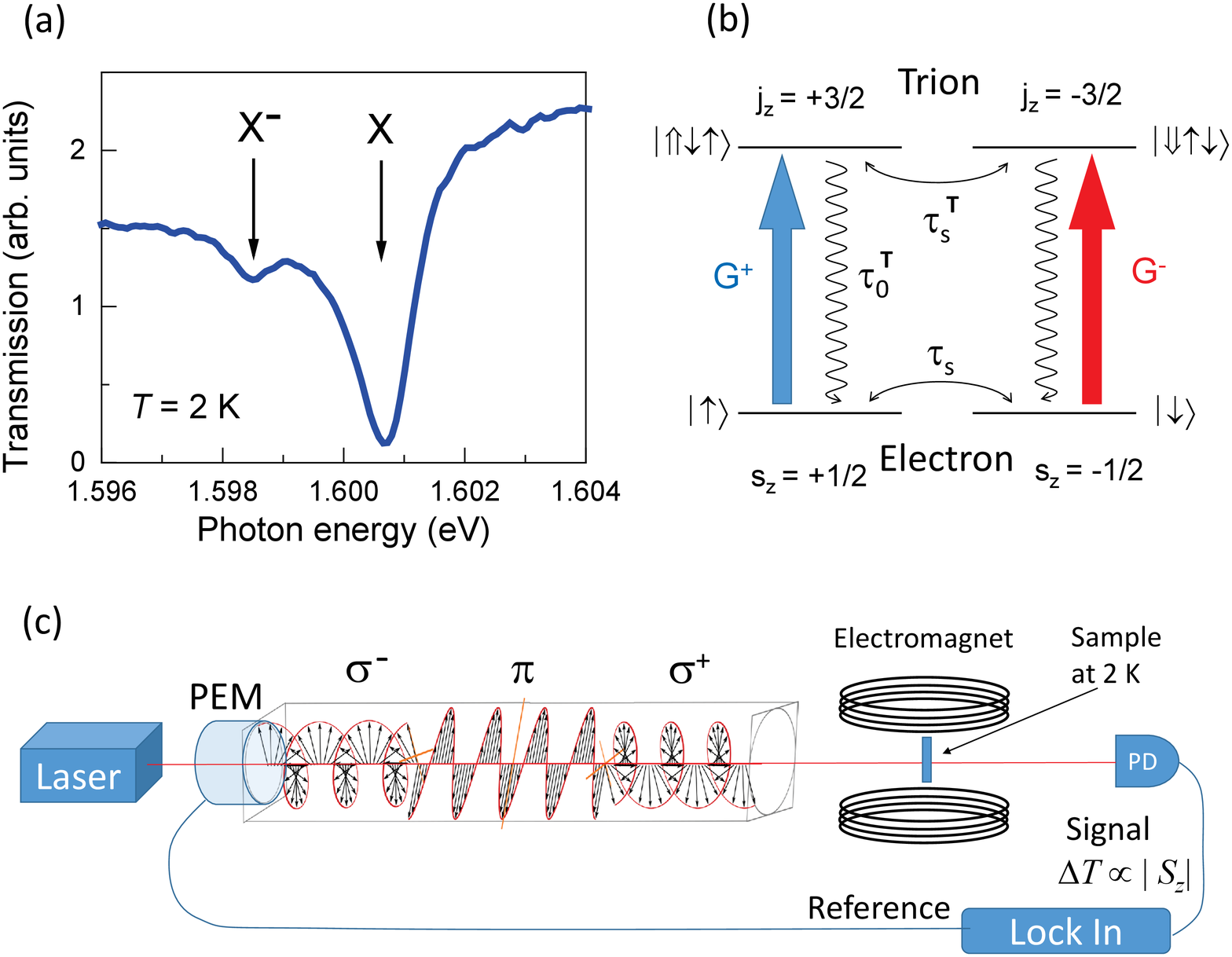}
\caption{(Color online) (a) Transmission spectrum of the studied n-doped CdTe/Cd$_{0.78}$Mg$_{0.22}$Te QWs structure measured with a white light source and a spectrometer equipped with a multichannel detector. (b) Energy level diagram including optical transitions for localized trion. (c) Scheme of the experimental setup. PEM and PD denote photo-elastic modulator and photodiode, respectively.}
\label{fig:schema}
\end{figure*}

In this work we demonstrate an alternative approach for the investigation of spin dynamics in semiconductors based on resonant optical pumping of resident electrons. In detail, a single laser beam tuned in resonance with the charged exciton (trion) transition in a semiconductor quantum well is used for spin pumping of resident electrons in their ground state. The absorption of the very same laser beam is used to monitor the electron spin polarization. The application of an external magnetic field in Voigt geometry leads to depolarization of the electron spin ensemble so that the Hanle effect is observed. Interestingly, Hanle peaks appear not only at $B=0$, but also around magnetic fields corresponding to the resonance condition $\Omega_L=2 \pi n f$ ($n$ is an integer), when a pulsed laser source with a relatively high repetition rate ($f \gg \tau_s^{-1}$) is used. The observations can be treated as optical pumping in the rotating frame, where the frequency of rotation corresponds to $f$ or higher harmonics thereof. Our approach resembles experiments performed by Bell and Bloom in atomic systems~\cite{Bell-Bloom-61}. Surprisingly, this technique has not yet been applied to semiconductors with relatively short spin relaxation times, as it requires laser sources with high repetition rates in the sub-GHz or even GHz range.

\section{Investigated sample}
\label{sec:samples}

The studied sample (reference number \#031901B) was grown by molecular-beam epitaxy on a (001) GaAs substrate with a Cd$_{0.78}$Mg$_{0.22}$Te buffer of 4.4~$\mu$m thickness. The intrinsic heterostructure comprises five 20~nm thick CdTe quantum well (QW) layers separated by 105~nm wide Cd$_{0.78}$Mg$_{0.22}$Te barriers. The barriers are thick enough to exclude electronic coupling between the QWs. The resident electrons in the QWs are provided by iodine donors in 15~nm wide barrier regions which are located at a distance of 20~nm from the QWs. The structure sequence was covered by a 110~nm Cd$_{0.78}$Mg$_{0.22}$Te cap layer. The sample was mounted on a sapphire disk and the GaAs substrate was removed chemically in order to perform transmission experiments.

The transmission spectrum of the sample at $T=2$~K is presented in Fig.~\ref{fig:schema}(a). The two dips at 1.5985~eV and 1.6007~eV are attributed to the resonance energies of the trion (X$^-$) and exciton (X) complexes, respectively. From the amplitude ratio between the exciton and trion features  we estimate the density of conduction band electrons in the QW to be $n_e\approx 10^{10}$~cm$^{-2}$~\cite{Astakhov-02}. For that small density we can assume that at low temperatures each resident electron is spatially isolated and localized in one of the potential fluctuations arising from variations of composition and width of the QW. The energy level structure and the possible optical transitions between the levels are shown in Fig.~\ref{fig:schema}(b). The ground state is a doublet with electron spin $s = 1/2$. The optically excited trion complex comprises two electrons and one hole. The X$^-$ in the singlet state is composed of two conduction band electrons with antiparallel spin (total spin $s=0$) and the hole with possible angular momentum projections $j_z=\pm3/2$, where the $z$-axis is defined by the QW confinement axis (growth direction of the structure) and parallel to the light propagation direction.

\section{Single beam approach}

The generation of trions through resonant excitation depends on the light polarization and the spin polarization of the resident electron ensemble. Only spin up electrons $|\uparrow\rangle$ are selectively excited through pumping with $\sigma^+$ polarized photons. The excited spin up  X$^-$  $|\Uparrow\uparrow\downarrow\rangle$ can subsequently decay either radiatively through the same channel or, after a hole spin-flip process to $|\Downarrow\uparrow\downarrow\rangle$, by emitting a $\sigma^-$ photon. The latter is responsible for creating spin down electrons. Hence, for repetitive resonant excitation of trions with $\sigma^+$ polarized light, a non-equilibrium state with more spin down than spin up electrons, i.e. optical pumping, is established~\cite{Akimov-07, Astakhov-08}. In turn, the optical pumping leads to a decrease of light absorption because the electron population of the ground state with spin projection $+1/2$ is depleted for $\sigma^+$ excitation and consequently the trion excitation rate decreases. Therefore, the spin polarization of resident electrons can be detected through bleaching of the trion absorption, which is manifested by a increase of the intensity of the transmitted light \cite{Atoms-bleaching}.

In an external magnetic field $\mathbf B$ oriented along the $x$-axis the equation of motion for the spin density of the resident electrons $\mathbf S$ reads
\begin{equation}
\label{Bloch-eq}
\frac{d\mathbf{S}}{dt} = \mathbf{P} - \frac{\mathbf S}{\tau_s} + \mathbf{\Omega \times S},
\end{equation}
where $\mathbf P = (0,0,P)$ is the spin pumping term, $\tau_s$ is the spin relaxation time of the resident electrons, and $\mathbf{\Omega} = (\Omega_L,0,0)$ defines the Larmor precession of $\mathbf S$ around the external magnetic field ~\cite{Akimov-07,Astakhov-08}. The spin pumping term depends on the helicity of the pump light
\begin{equation}
\label{P-term}
P=-\frac{ n_e \rho_c G}{2}\frac{\tau_0^T}{\tau_0^T+\tau_s^T},
\end{equation}
where $G = (G^+ +G^-)/2$ is the optical generation rate, which is proportional to the laser intensity $I$, $\rho_c=(G^+-G^-)/(G^++G^-)$ is the circular polarization degree of the exciting light, $\tau_0^T$ is the trion lifetime, and $\tau_s^T$ is the trion spin relaxation time. The rates $G^+$ and $G^-$ correspond to the contributions from the $\sigma^+$ and $\sigma^-$ polarized light components. Equation \ref{Bloch-eq} is valid for low pumping rates $P\tau_s\ll n_e/2$, i.e. when the spin polarization is small and no saturation effects are observed~\cite{Astakhov-08}. In addition, we assume that $\tau_s \gg \tau_0^T, \tau_s^T$. This condition is satisfied for the investigated CdTe/(Cd,Mg)Te QW structure for which $\tau_0^T\approx 50$~ps and $\tau_s^T=1000$~ps~\cite{Zhukov-07, Langer-14}.

For a low pumping rate, the population of trions is significantly smaller than the population of the ground state and the absorption of light is given by
\begin{equation}
\label{eq:A}
A \propto \left( \frac{n_e}{2} + \rho_c S_z \right) I.
\end{equation}
The second term on the right hand side of Eq.~\eqref{eq:A} determines the spin dependent part of the absorption due to optical pumping. For $\sigma^+$ polarized light ($\rho_c>0$), the optical pumping leads to negative spin polarization $S_z<0$, while in the opposite case $\sigma^-$ excitation ($\rho_c<0$) leads to $S_z>0$. Therefore, for circularly polarized light the transmitted light intensity is always increased due to bleaching of the absorption, while for linearly polarized excitation there is no optical pumping ($S_z=0$) and the transmitted intensity is minimum. Note that the changes in transmission are determined by the degree of circular polarization of the incident laser light but do not depend on the sign of polarization. The difference in the transmitted light intensity, $\Delta T$, for circularly ($\rho_c=\pm1$) and linearly polarized ($\rho_c=0$) light is given by: $\Delta T \propto |S_z| I$.

The steady state solution of Eq.~\eqref{Bloch-eq} for continuous wave excitation corresponds to the Hanle curve
\begin{equation}
\label{Hanle-eq}
S_{z0} = \frac{ P \tau_s }{ 1 + \Omega_L^2 \tau_s^2 }.
\end{equation}
For excitation with a pulsed laser source the excitation rate becomes time dependent $G(t) = G_0 \sum^{+\infty}_{-\infty} \exp{[i2\pi n f t]}$, where $G_0$ is proportional to the time-integrated laser intensity $I$. Using Eq.~\eqref{Bloch-eq} we obtain the steady state solution
\begin{equation}
\label{eq:Signal}
\frac{\Delta T}{I} \propto \sum^{+\infty}_{-\infty}|S_{zn}|,
\end{equation}
where $S_{z0}$ is given by Eq.~\eqref{Hanle-eq} and corresponds to the Hanle curve around $B=0$, while the higher harmonics with $|n|>0$ correspond to optical pumping of the resident electron spins in the rotating frame
\begin{equation}
\label{eq:Rot-S}
S_{zn}=\frac{1}{2}S_{z0}(\Omega_L-2\pi n f).
\end{equation}
The factor $1/2$ on the right hand in Eq.~\eqref{eq:Rot-S} side appears in the rotating frame. This is because the harmonic excitation signal can be expressed as sum of two terms with half the amplitude that rotate in opposite directions. In the rotating frame only one of these terms contributes to the steady state solution.  The harmonics with $n\neq 0$ correspond to cloning of the Hanle effect, when the Larmor precession frequency $\Omega_L$ is replaced by $\Omega_L-2\pi n f$ in Eq.~\eqref{Hanle-eq}.

\section{Experimental realization}
\label{sec:exp}

The experiments were performed using three different tunable Ti:Saphir lasers: (i) a continuous wave (cw) laser (T\&D Scan, Tekhnoscan); (ii) a self mode-locked picosecond oscillator (Mira 900D, Coherent) with a repetition rate of $f=75.8$~MHz, a spectral width of 0.9~nm and a pulse duration of about 2~ps; (iii) a self mode-locked femtosecond oscillator (Gigajet 20c, Laser Quantum) with $f=999$~MHz, a spectral width of 30~nm and a pulse duration of 50~fs. The latter laser was used in combination with a pulse shaper which reduces the spectral width of the pulses down to 0.7~nm and correspondingly increases the duration to several ps. The sample was placed in a liquid helium bath cryostat with a variable temperature inset and kept at a temperature of $T=2$~K. An external magnetic field $B$ of up to 700~mT was applied by an electromagnet in the Voigt geometry, e.g. parallel to the quantum well plane. The photon energy of the exciting light was tuned to the trion resonance and the transmitted light intensity was measured with a photodiode (see Fig.~\ref{fig:schema}(c)).

\begin{figure}[htpb]
\centering
\includegraphics[width=\linewidth]{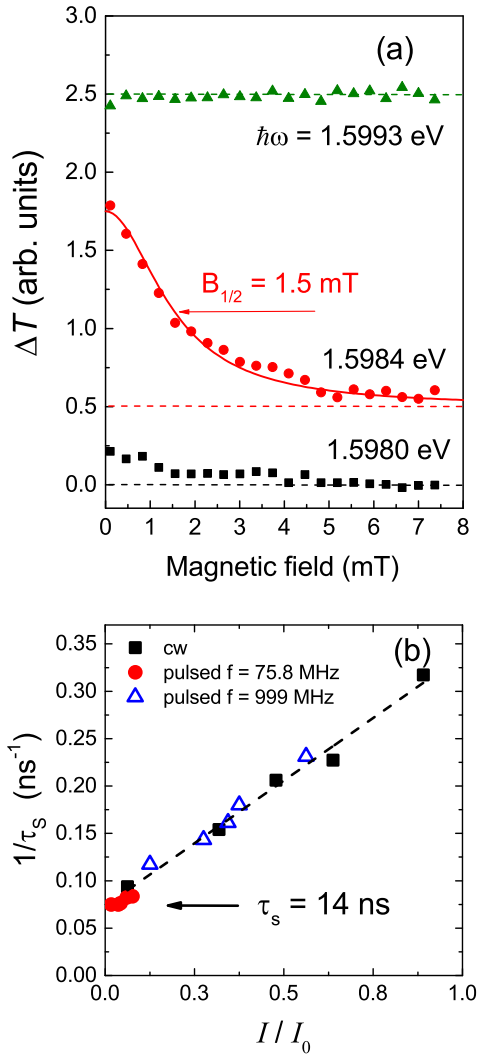}
\caption{(Color online) (a) Magnetic field dependence of the differential transmission $\Delta T$ measured with cw excitation for different photon energies $\hbar\omega$ and excitation intensity $I \approx 2  $~W/cm$^2$. The solid curve corresponds to a Lorentz fit after Eq.~\eqref{Hanle-eq}. The curves are shifted vertically for clarity. The dotted lines correspond to the zero signal level. (b) Power dependence of the spin relaxation rate evaluated with Eq.~\eqref{Hanle-eq} under excitation of the trion at photon energy $\hbar\omega=1.5984$~eV for the cw laser and the pulsed laser sources with $f=75.8$ and 999~MHz. Intensity $I_0 \approx $ 4~W/cm$^2$. $T=2$~K.}
\label{Fig:cw}
\end{figure}

In order to detect the difference in absorption caused by the optical pumping of resident electron spins, the laser light was modulated between linear and circular polarization with a photo-elastic modulator (PEM) at frequency $\nu\approx50$~kHz (see Fig~\ref{fig:schema}(c)). The PEM modulates the phase between the two orthogonal linearly polarized components in time $t$, $\phi=\frac{\pi}{2}\sin(2 \pi \nu t)$. In this case the polarization state of the laser beam changes from $\pi$ (linear) to $\sigma^+$ and back to $\pi$ within the first half of the period, and the polarization sequence is $\pi - \sigma^- - \pi$ within the second half-period of modulation. The differential intensity signal $\Delta T$ was demodulated by a lock-in amplifier at double frequency $2\nu$. The modulation of the light polarization with the PEM is slow compared to the spin dynamics in the studied system. Therefore, the spin polarization in the ground state adiabatically follows the polarization of light, i.e. according to Eq.~\eqref{Hanle-eq}, optical pumping is present for $\sigma$ polarized excitation while it is absent for $\pi$ polarized light.

\section{Demonstration of single beam detection}
\label{sec:demo}

\subsection{CW excitation}

The magnetic field dependencies of the differential transmission $\Delta T$ measured with cw excitation for different photon energies $\hbar\omega$ are shown in Fig.~\ref{Fig:cw}(a). Changes in transmission are present only for resonant excitation of trion ($\hbar\omega= 1.5984$~eV), i.e. one has to excite the resident carriers to achieve optical pumping and detection of their spin polarization. The magnetic field dependence of $\Delta T$ is well described by a Hanle curve following Eq.~\eqref{Hanle-eq} with a halfwidth at half maximum $B_{1/2}$, which increases with increasing laser intensity. The corresponding spin relaxation time $\tau_s$ varies from 11~ns to 4~ns. The dependence of $\tau_s^{-1}$ on excitation power $I$ is presented in Fig.~\ref{Fig:cw}(b). It increases linearly with $I$, which can be attributed due to a number of reasons. First, the increase of the spin relaxation rate with increasing pump density can be understood in terms of a delocalization of the resident electrons, caused by their heating due to interaction with photogenerated carriers~\cite{Zhukov-07}. Second, local recharging processes such as the recombination with holes limit the lifetime of the resident electrons and influence their degree of spin polarization ~\cite{Dzhioev-02, Akimov-07}. Finally, at larger excitation intensities the optical pumping enters the non-linear regime and the evaluation of the spin relaxation with Eq.~\eqref{Hanle-eq} is not valid. Nevertheless, independent of the particular mechanism it is possible to determine the true spin relaxation time in the limit of low pump intensities. From the linear fit of the power dependence in Fig.~\ref{Fig:cw}(b) we evaluate $\tau_s = 14$~ns for the studied structure.

\begin{figure}[htb]
\includegraphics[width=\linewidth]{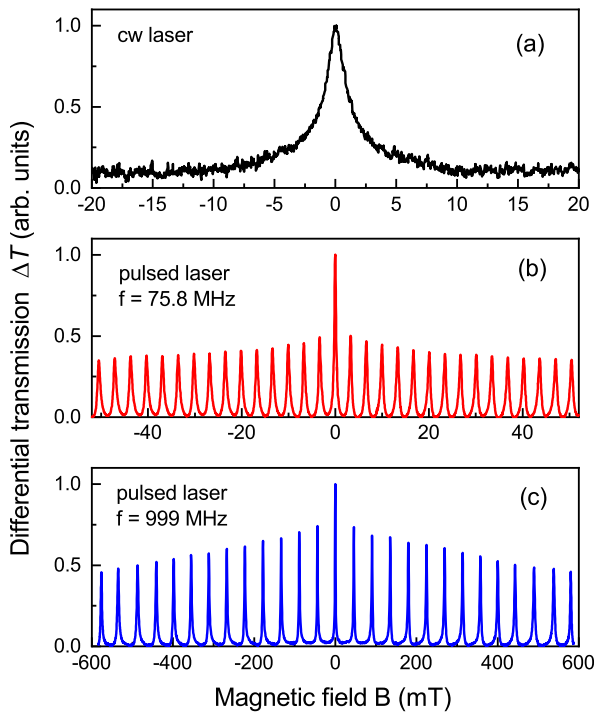}
\caption[width=\textwidth]{(Color online) Magnetic field dependence of the differential transmission for resonant excitation of the trion ($\hbar\omega = 1.5984$~eV) with the cw (a) and the pulsed lasers with $f = 75.8$~MHz (b) and $f=999$~MHz (c). $T=2$~K.}
\label{Fig:RSA}
\end{figure}

\subsection{Comparison with pulsed excitation}

The Hanle curves measured for cw and pulsed excitation are summarized in Fig.~\ref{Fig:RSA}. While for cw pumping we observe a single Hanle peak at $B=0$, excitation with a periodic sequence of pulses leads to a large set of Hanle peaks, which appear every $3.36$ and 44.2~mT for $f=75.8$ and 999~MHz, respectively. In contrast to cw excitation, here we are able to evaluate precise information on the Larmor precession frequency $\Omega_L$. From the resonance condition for optical orientation in the rotating frame, $g \mu_B B = 2 \pi n \hbar f $, we evaluate $|g| = 1.61$. This value is in agreement with the $g$ factor of resident electrons in similar QW structures as evaluated from oscillatory pump-probe signals~\cite{Zhukov-07}. The widths of the Hanle peaks for different harmonics $n$ are close to each other in weak magnetic fields up to 100~mT. For larger magnetic fields the Hanle peaks become broader due to the faster dephasing of the spin ensemble as a consequence of the inhomogeneous broadening of the g-factor~\cite{Zhukov-07}. The power dependence of the relaxation rate evaluated for $B<100$~mT is shown for the different excitation lasers in Fig.~\ref{Fig:cw}(b). Similarly to cw laser excitation, it also depends linearly on power and approaches the true spin relaxation time $\tau_s=14$~ns in the limit of low pumping rates. Thus, the use of pulsed excitation allows us to obtain the desired information on both the $g$ factor and relaxation dynamics. Knowledge of the $g$ factor is important for a proper interpretation of the spin dynamics because its value allows determining the type of carriers which are spin polarized, i.e., holes or electrons.

\section{Conclusions}

In conclusion, we have demonstrated a new experimental approach for investigation of the spin dynamics in semiconductor nanostructures. It is based on optical pumping of resident electron spins in the rotating frame using a periodic sequence of laser pulses with a high repetition rate and simultaneously measuring the transmitted intensity which monitors the spin polarization of the electron ensemble. In contrast to cw excitation, the Hanle effect appears at every magnetic field at which the Larmor precession frequency is synchronized with the pulse repetition rate. This allows one to determine the key parameters of the spin dynamics, i.e. the $g$ factor of the optically pumped carriers and their spin relaxation time. The sequence of multiple Hanle curves can be used to evaluate the spin relaxation times across a large range of magnetic fields. In particular, this relatively simple approach can be efficiently used for the investigation of dynamic nuclear polarization which changes the Larmor precession frequency of conduction band electrons.

\section*{Acknowledgements}
We acknowledge the financial support by the Deutsche Forschungsgemeinschaft through the International Collaborative Research Centre 160 (Projects A3 and C2), the Russian Foundation for Basic Reasearch (Project N 15-52-12017 NNIO\_a) and the DAAD (project EXCIPLAS).  The research in Poland was partially supported by the National Science Centre (Poland) through Grants No. DEC-2012/06/A/ST3/00247 and No. DEC-2014/14/M/ST3/00484, as well as by the Foundation for Polish Science through the IRA Programme co-financed by EU within SG OP.

\end{document}